\documentclass{article} 
\usepackage{iclr2021_conference,times}

\usepackage{amsmath,amsfonts,bm}

\def\eqref#1{equation~\ref{#1}}

\def\1{\bm{1}}

\DeclareMathAlphabet{\mathsfit}{\encodingdefault}{\sfdefault}{m}{sl}
\SetMathAlphabet{\mathsfit}{bold}{\encodingdefault}{\sfdefault}{bx}{n}

\usepackage{hyperref}
\usepackage{url}

\usepackage{graphbox}  

\newif\ifdraft\draftfalse

\ifdraft
\usepackage[draft]{commenting}
\else
\usepackage{commenting}
\fi

\declareauthor{ag}{ag}{orange}
\authorcommand{ag}{comment}
\declareauthor{ed}{ed}{blue}
\authorcommand{ed}{comment}
\declareauthor{ma}{ma}{green}
\authorcommand{ma}{comment}
\declareauthor{ad}{ad}{purple}
\authorcommand{ad}{comment}
\declareauthor{yw}{yw}{red}
\authorcommand{yw}{comment}

\title{COIN: COmpression with Implicit Neural\\ representations}

\author{Emilien Dupont*, Adam Goliński*, Milad Alizadeh, Yee Whye Teh \& Arnaud Doucet\\
University of Oxford\\
\texttt{dupont@stats.ox.ac.uk,adamg@robots.ox.ac.uk}
}

\iclrfinalcopy 
\begin{document}

\maketitle

\begin{abstract}
We propose a new simple approach for image compression: instead of storing the RGB values for each pixel of an image, we store the weights of a neural network overfitted to the image. 
Specifically, to encode an image, we fit it with an MLP which maps pixel locations to RGB values. 
We then quantize and store the weights of this MLP as a code for the image. 
To decode the image, we simply evaluate the MLP at every pixel location. 
We found that this simple approach outperforms JPEG at low bit-rates, \textit{even without entropy coding or learning a distribution over weights}. 
While our framework is not yet competitive with state of the art compression methods, we show that it has various attractive properties which could make it a viable alternative to other neural data compression approaches.
\end{abstract}

\section{Introduction}

\begin{figure}[b!]
\begin{center}
\includegraphics[height=70pt, align=c]{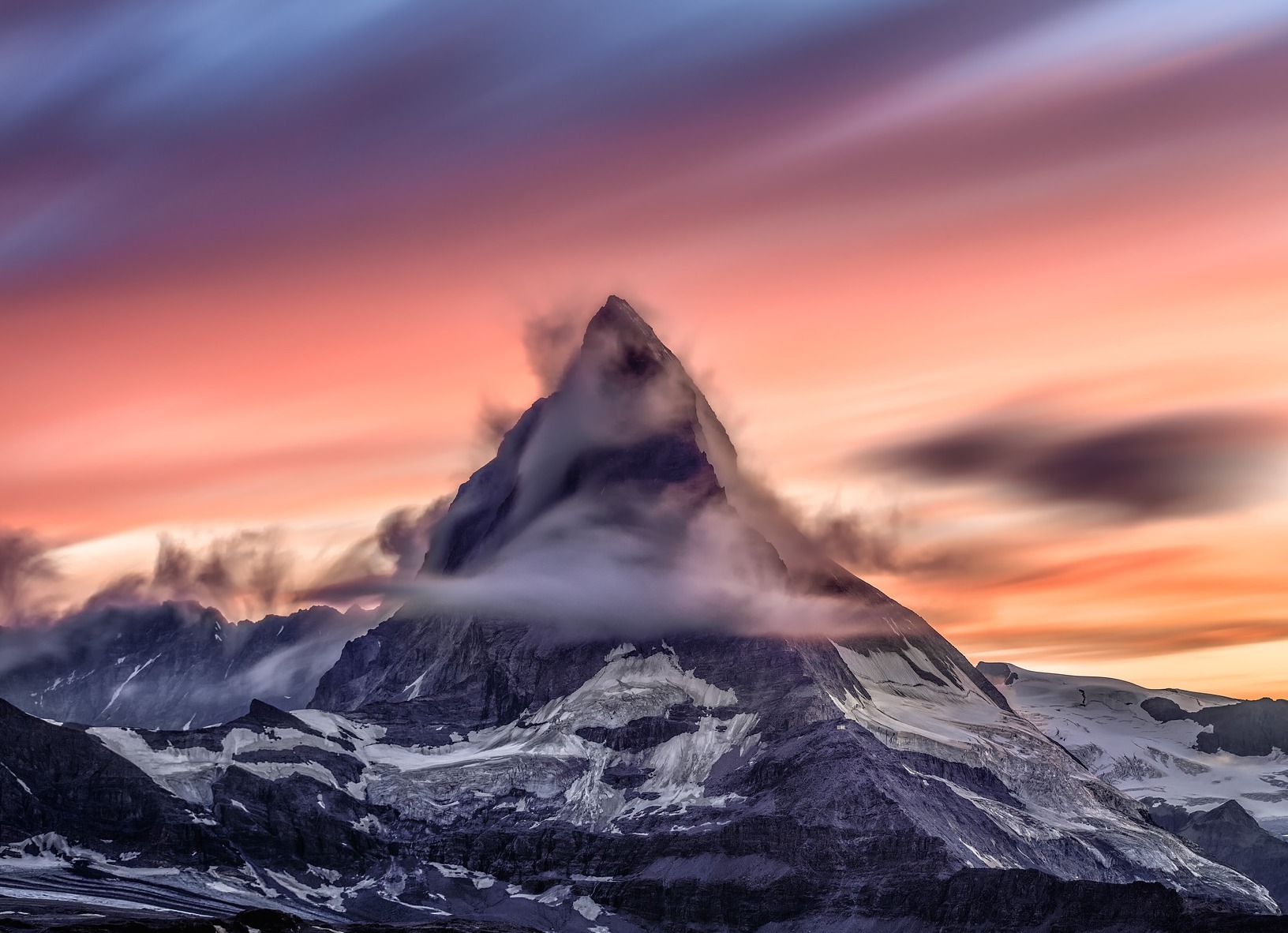}
\hspace{5pt} {\LARGE $\to$} \hspace{5pt} 
\raisebox{2pt}{\large $\begin{matrix}x\\y\end{matrix}$} 
{\LARGE $\underbrace{\includegraphics[height=65pt, align=c]{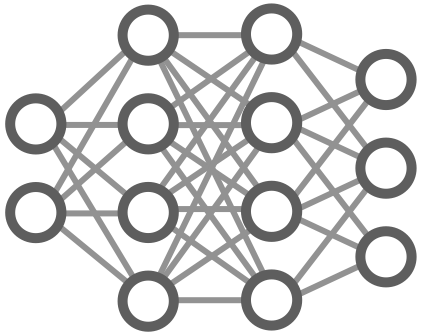}}_{\theta}$} 
\raisebox{0pt}{\large $\begin{matrix}R\\[2pt]G\\[2pt]B\end{matrix}$} 
\hspace{5pt} {\LARGE $\to$} \hspace{5pt} 
{\LARGE $\lfloor \theta \rfloor$}
\hspace{5pt} {\LARGE $\to$} \hspace{5pt} 
\includegraphics[height=50pt, align=c]{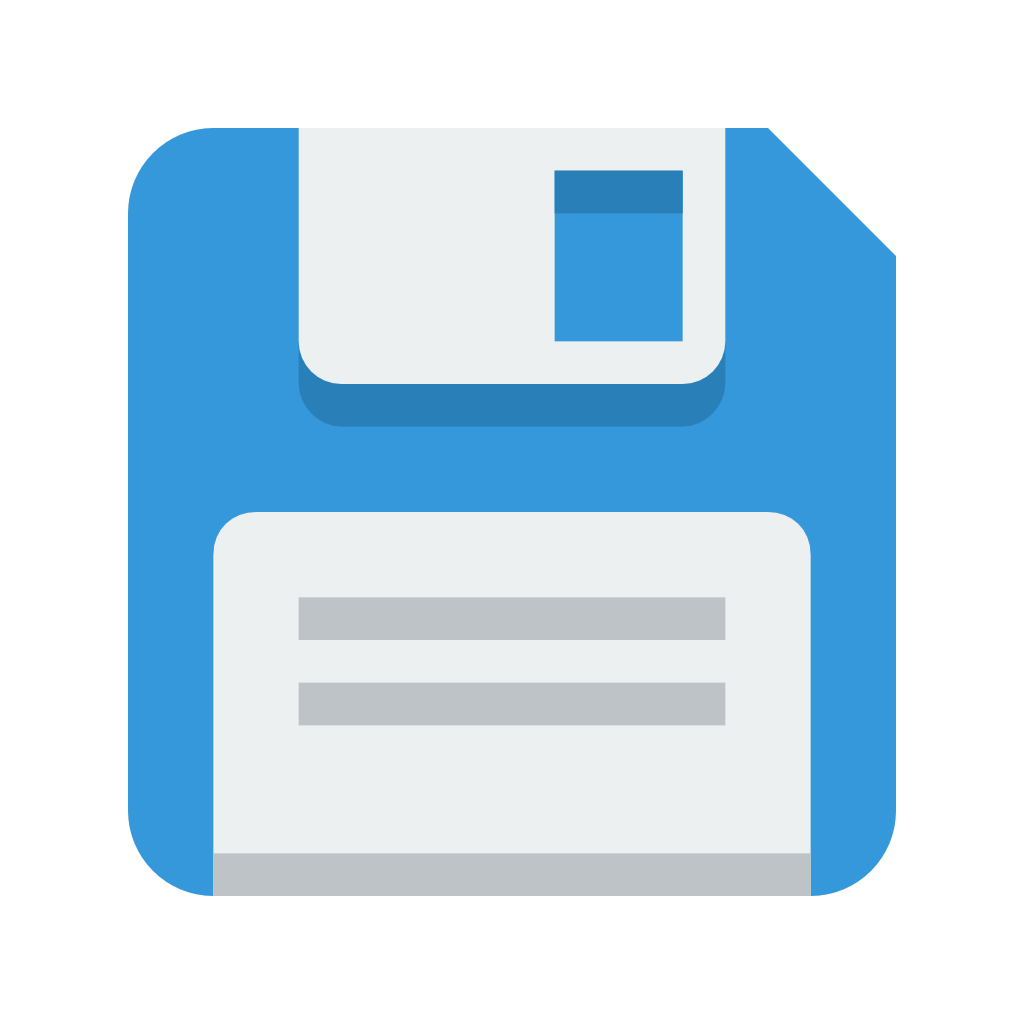}
\end{center}
\vspace{-10pt}
\caption{Compressed implicit neural representations. We overfit an image with a neural network mapping pixel locations $(x, y)$ to RGB values (often referred to as an implicit neural representation). We then quantize the weights $\theta$ of this neural network to a lower bit-width and transmit them.}\label{fig-intro}
\end{figure}


Neural image compression methods typically operate in an autoencoder setup \citep{balle2018variational, minnen2018joint, lee2019contextadaptive}. 
The sender uses an encoder to map the input data to a discretized latent code, which is then entropy coded into a bitstream according to a learned latent distribution.
The bitstream is transmitted to the receiver that decodes it into a latent code, which is finally passed through the decoder to reconstruct the image.

In this paper, we take a different approach: we encode an image by overfitting it with a small MLP mapping pixel locations to RGB values and then transmit the weights $\theta$ of this MLP as a code for the image (see Figure \ref{fig-intro}). While overfitting such MLPs, referred to as implicit neural representations, is difficult due to the high frequency information contained in natural images \citep{basri2019convergence, tancik2020fourier}, recent research has shown that this can be mitigated by using sinusoidal encodings and activations \citep{mildenhall2020nerf, tancik2020fourier, sitzmann2020implicit}. In this work, we show that using MLPs with sine activations, often referred to as SIRENs \citep{sitzmann2020implicit}, we can fit large images (393k pixels) with surprisingly small networks (8k parameters). 

We evaluate our method on standard image compression tasks and show that we outperform JPEG at low bit-rates, even without entropy coding or learning a distribution over weights. 
As implicit representations have successfully been applied in the context of generative modeling \citep{dupont2021generative}, 
it is likely that combining our approach with a learned weight distribution could lead to promising new approaches for neural data compression.
Further, by treating our image as a function from pixel locations to RGB values, we can perform progressive decoding simply by evaluating our function at progressively higher resolutions, which is particularly attractive for resource constrained receiving devices.

\section{Method}
In this section, we describe COmpressed Implicit Neural representations (COIN), our proposed method for image compression. The encoding step consists in overfitting an MLP to the image, quantizing its weights and transmitting these.
At decoding time, the transmitted MLP is evaluated at all pixel locations to reconstruct the image.

\subsection{Encoding}
\label{sec:encoding}
Let $I$ denote the image we wish to encode, such that $I[x, y]$ returns the RGB values at pixel location $(x, y)$. We define a function $f_\theta : \mathbb{R}^2 \to \mathbb{R}^3$ with parameters $\theta$ mapping pixel locations to RGB values in the image, i.e., $f_\theta(x,y) = (r, g, b)$. We can then encode the image by overfitting $f_\theta$ to the image under some distortion measure. In this paper, we use the mean squared error, resulting in the following optimization problem
\begin{equation}\label{eq-distortion}
    \min_{\theta} \sum_{x,y} \| f_\theta(x,y) - I[x, y] \|_2^2, 
\end{equation}
where the sum is over all pixel locations.

Choosing the parameterization of $f_\theta$ is crucial. Indeed, parameterizing $f_\theta$ by an MLP with standard activation functions results in underfitting, even when using a large number of parameters \citep{tancik2020fourier, sitzmann2020implicit}. 
This problem can be overcome in multiple ways, e.g. 
by encoding pixel coordinates with Fourier features \citep{tancik2020fourier} or by using sine activation functions \citep{sitzmann2020implicit}. 
Empirically, we found that the latter option 
yielded 
better
results for a given parameter budget.

Minimizing equation (\ref{eq-distortion}) is trivial given a large enough MLP. 
However, we store the parameters $\theta$ of the MLP as the compressed 
description of the image, 
so restricting the number of weights will improve the compression rate. 
The goal is therefore to fit $f_\theta$ to $I$ (i.e., minimize distortion) using the fewest parameters possible (i.e., maximizing rate). 
Our approach then effectively converts a data compression problem to a model compression problem.

To reduce the model size, we consider two approaches: architecture search and weight quantization. More specifically, we perform a hyperparameter sweep over the width and number of layers of the MLP and quantize the weights from 32-bit to 16-bit precision, which was sufficient to outperform the JPEG standard for low bit-rates. However, we believe that more sophisticated approaches to architecture search \citep{elsken2019neural} and especially model compression \citep{ullrich2017soft,havasi2019minimal,vanbaalen2020bayesian} 
will further improve results.


\ag{@Milad: verify the citations above are all relevant please and add others you think are relevant, especially about straight up quantization rather than Bayesian approaches.
Any other thoughts?
The same comment is at the end of Related Work section.} 

\subsection{Decoding}
\label{sec:decoding}
Given the stored quantized weights $\theta$, decoding simply consists in evaluating the function $f_\theta$ at every pixel location to reconstruct the image.
This decoding approach gives us extra flexibility: we can progressively decode the image, e.g. by decoding parts of the image or a low resolution image first, simply by evaluating the function at various pixel locations. Partially decoding images in this way is difficult with autoencoder based methods, showing a further advantage of the COIN approach.

\begin{figure}[t]
    \centering
    \begin{minipage}{0.65\textwidth}
        \begin{center}
        \includegraphics[height=160pt]{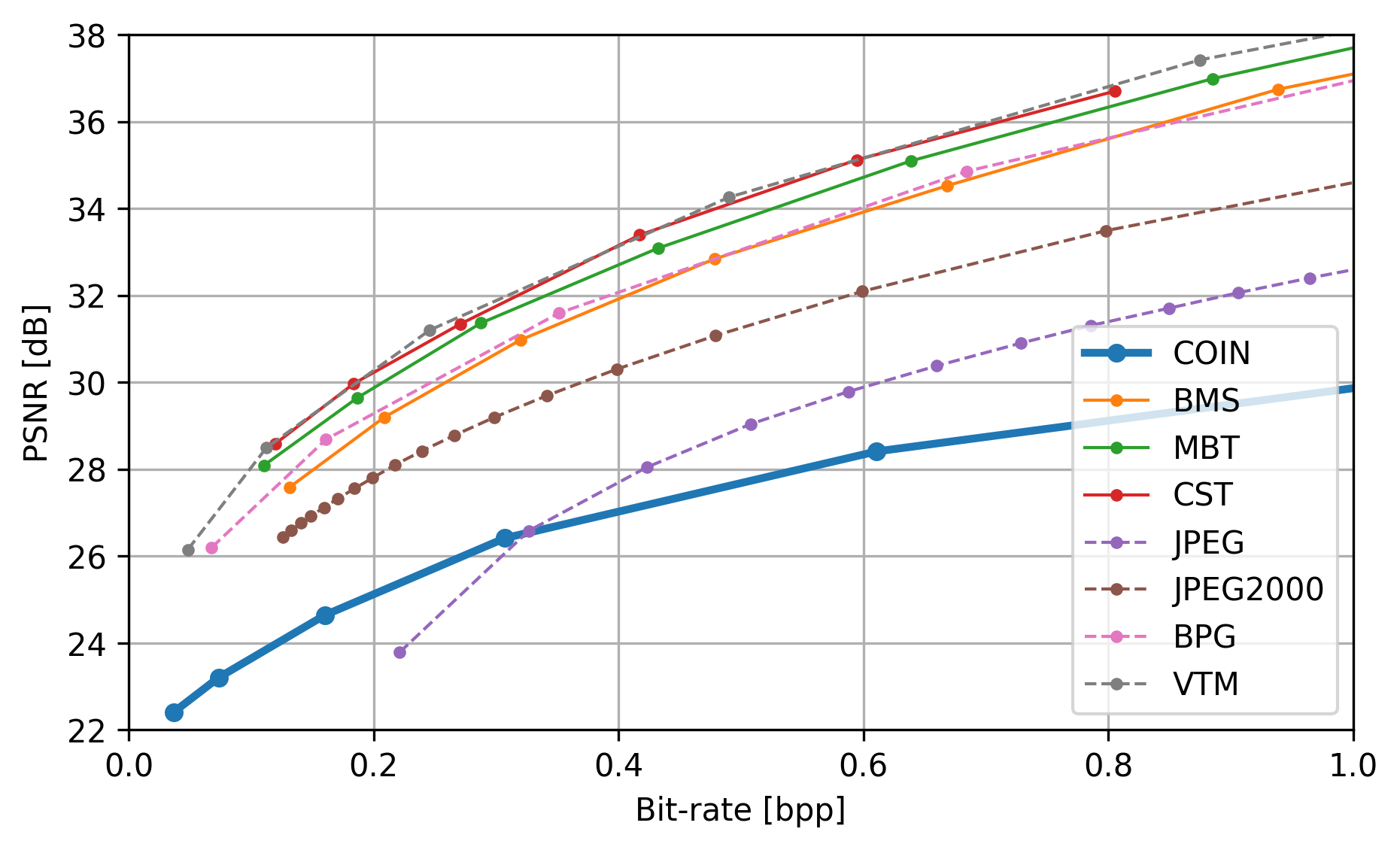} 
        \end{center}
        \vspace{-10pt}
        \caption{Rate distortion plots on the Kodak dataset.} \label{fig-rate-dist-kodak}  
    \end{minipage}\hfill
    \begin{minipage}{0.35\textwidth}
        \begin{center}
        \includegraphics[height=150pt]{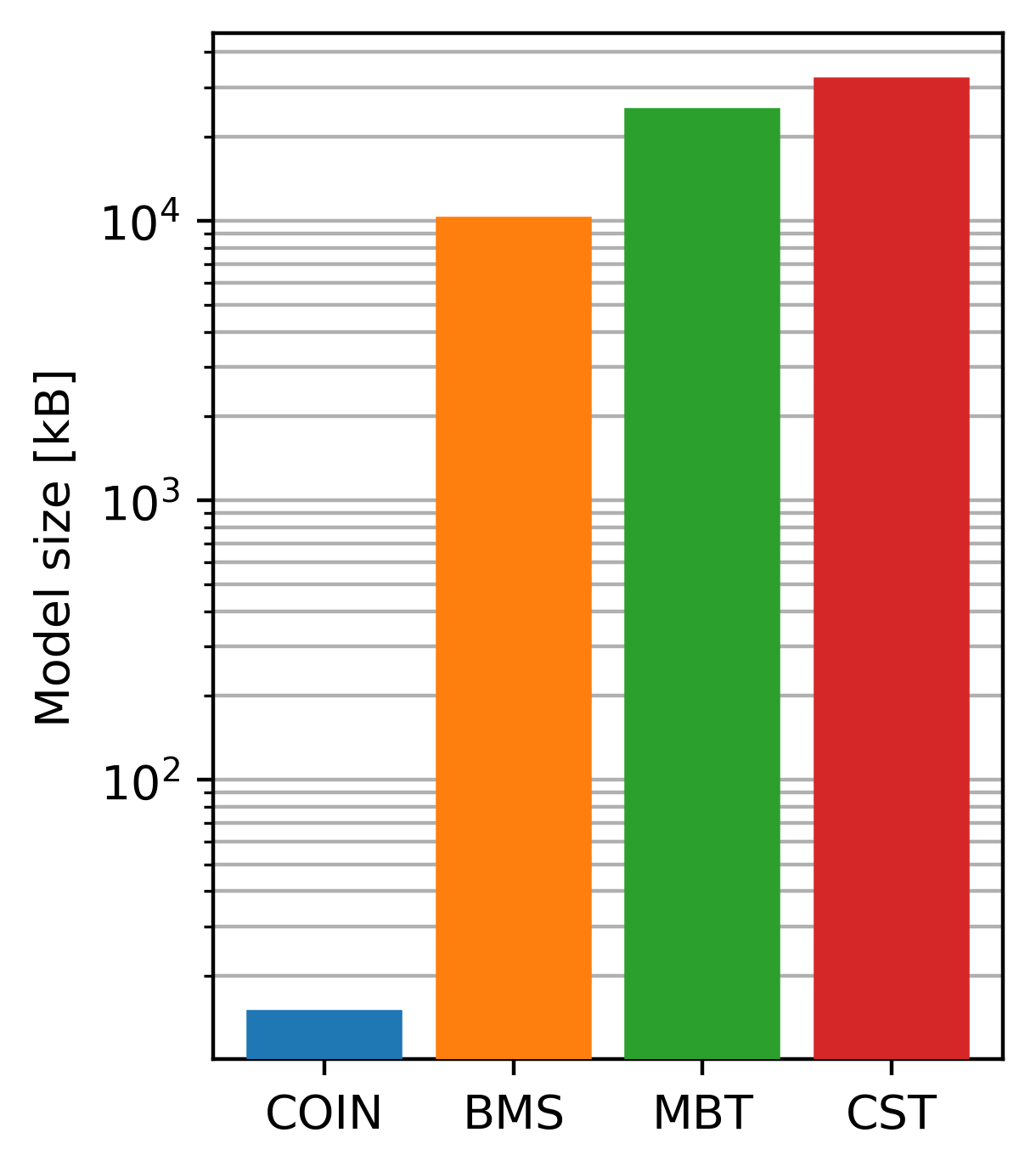}
        \end{center}
        \caption{Model sizes at 0.3bpp.} \label{fig-model-sizes} 
    \end{minipage}
\end{figure}

\section{Related Work}

\textbf{Implicit neural representations}. Representing data with neural networks was originally proposed by \citep{stanley2007compositional} but has seen a recent surge in interest in the 3D vision community \citep{park2019deepsdf, niemeyer2019occupancy, chen2019learning}. Motivated by the fact that memory requirements of deep voxel representations grow cubically \citep{nguyen-phuoc2018rendernet, sitzmann2019deepvoxels, dupont2020enr}, implicit representations were proposed to compactly encode high resolution signals \citep{mildenhall2020nerf, tancik2020fourier, sitzmann2020implicit}. While the MLPs used to represent images typically have a relatively small number of parameters \citep{dupont2021generative}, we take this even further and show that by carefully choosing the architecture of the MLP and quantizing the weights, we can fit images with MLPs that take up significantly less space than storing RGB values. 



\textbf{Neural data compression}. 
Learned image compression methods are commonly based on hierarchical variational autoencoders \citep{balle2018variational,minnen2018joint,lee2019contextadaptive} with a learned prior and latent variables being discretized for the purpose of entropy coding. 
In a similar vein to work in the latent variable model literature \citep{hjelm2016iterative,krishnan2018challenges,kim2018semiamortized,marino2018iterative}, several works \citep{campos2019content,guo2020variable,yang2020improving} attempt to close the amortization gap \citep{cremer2018inference} by performing iterative gradient-based optimization steps on top of the use of amortized inference networks. 
\citep{yang2020improving} additionally identify and attempt to close the discretization gap stemming from the quantization of the latent variables, also by inference time per-instance optimization. \citep{vanrozendaal2021overfitting} take the idea of per-instance optimization of the model further: they perform per-instance finetuning of the decoder and transmit the quantized decoder parameter updates along with the latent code, leading to improved rate-distortion performance. 
In this paper, we take a different, and even more extreme from the per-instance optimization perspective, approach: we optimize an MLP to overfit a single image and transmit its weights as the compressed description of the image.

\textbf{Model compression}. 
While it is known that the problems of data and model compression are very closely related,
COIN explicitly casts the problem of data compression into a problem of model compression.
There exists a rich body of literature on model compression, \citep{ullrich2017soft,louizos2017bayesian,havasi2019minimal,vanbaalen2020bayesian,krishnamoorthi_whitepaper,
jacob2018quantization}, which could likely be used to improve COIN's performance.



\section{Experiments}

We perform experiments on the Kodak image dataset \citep{kodakdataset} consisting of 24 images of size $768\times512$. 
We compare our model against three autoencoder based neural compression baselines which we refer to as BMS \citep{balle2018variational}, MBT \citep{minnen2018joint}, and CST \citep{cheng2020learned}. 
We also compare against the JPEG, JPEG2000, BPG and VTM image codecs. 
To benchmark our model, we use the CompressAI library \citep{begaint2020compressai} and the pre-trained models provided therein. 
We implement our model in PyTorch \citep{paszke2019pytorch} and perform all experiments on a single RTX2080Ti GPU.

\textbf{Rate-distortion plots}. 
To determine the best model architectures for a given parameter budget (measured in bits per pixel or bpp\footnote{$\text{bits-per-pixel} = \frac{\text{\#parameters} \times \text{bits-per-parameter}}{\text{\#pixels}}$}), we first find valid combinations of depth and width for the MLPs representing an image. 
For example, for 0.3bpp using 16-bit weights, valid networks include MLPs with 10 layers of width 28, 7 layers of width 34 and so on.
We then select the best architecture by running a hyperparameter search over learning rates and valid architectures on a single image using Bayesian optimization (we found that the results of the architecture search transferred well to other images). The resulting model is trained on each image in the dataset at 32-bit precision and converted to 16-bit precision after training. We note that decreasing the weights' precision from 32-bit to 16-bit after training resulted in almost no distortion increase, but decreasing them further to 8-bit incurred significant amount of distortion, outweighing the benefit of halving the bpp.

Results of this procedure for various bpp levels are shown in Figure \ref{fig-rate-dist-kodak}. 
As can be seen, at low bit-rates our model improves upon JPEG \emph{even without using entropy coding}. 
While our approach is still far from the state of the art compression methods, we believe the performance of such a simple approach is promising for future work in this direction. 

\textbf{Model size}. In contrast to most other neural data compression algorithms, our method does not require a decoder at test time. 
Indeed, while the latent code representing the compressed image in such methods is small, the decoder model is large (typically much larger than an uncompressed image). 
As such, the memory required on the decoding device is also large. In our case, we only require the weights of a (very small) MLP on the decoder side, leading to memory requirements that are orders of magnitude smaller.
As can be seen in Figure \ref{fig-model-sizes}, at 0.3bpp, our method requires 14kB, whereas other baselines require between 10MB and 40MB. 



\begin{figure}[t]
    \centering
    \begin{minipage}{0.48\textwidth}
        \begin{center}
        \includegraphics[width=\textwidth]{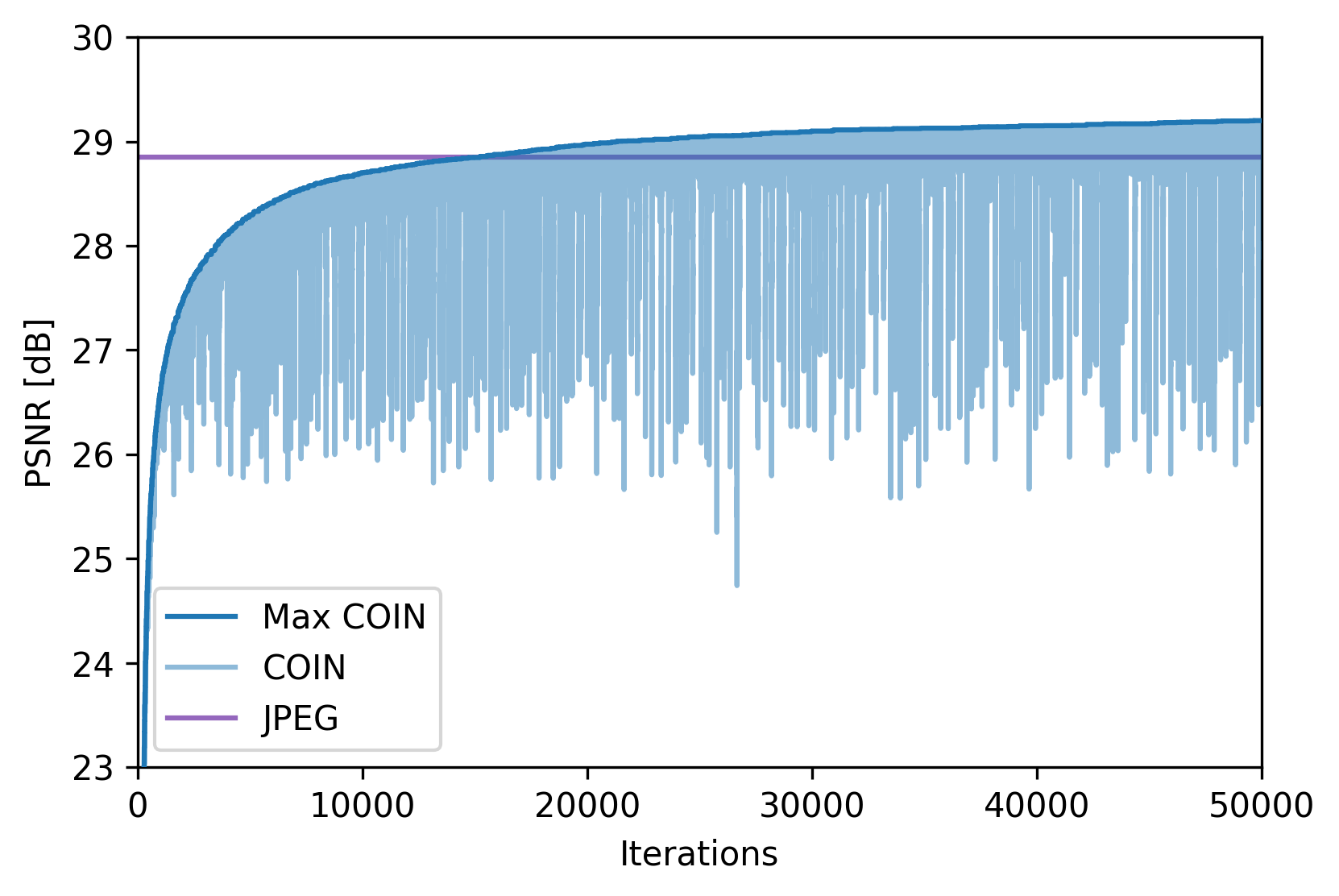}
        \end{center}
        \vspace{-15pt}
        \caption{Model training on image 15 in the Kodak dataset. Max COIN represents the max PSNR achieved at any point during training.} \label{fig-psnr-train}
    \end{minipage}\hfill
    \begin{minipage}{0.48\textwidth}
        \begin{center}
        \includegraphics[width=\textwidth]{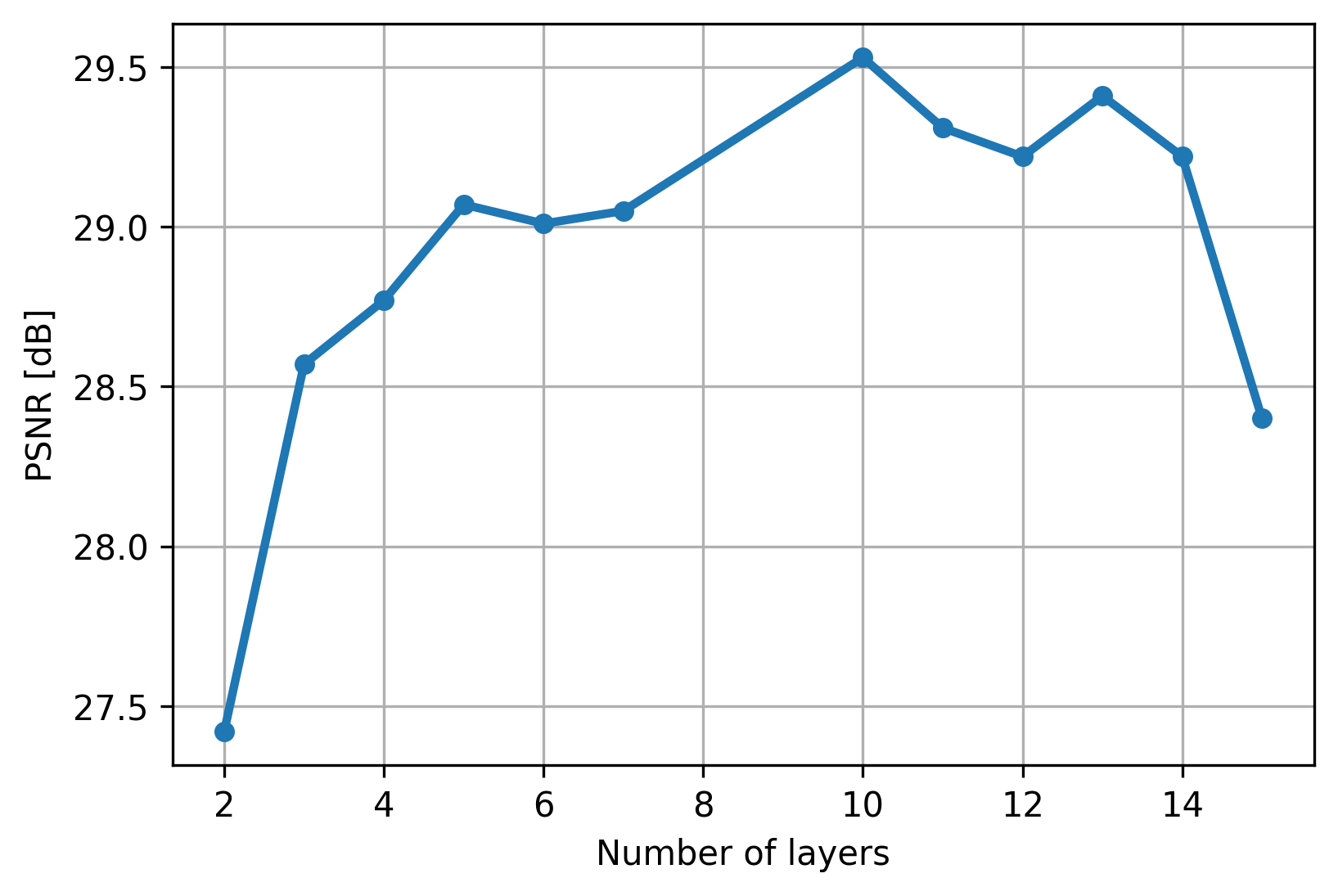}
        \end{center}
        \vspace{-15pt}
        \caption{Plot of maximum PSNR for networks of the same size (0.3bpp) with different architectures.} \label{fig-arch-choice}
    \end{minipage}
\vspace{-10pt}
\end{figure}

\textbf{Encoding optimization dynamics}. We show an example of the overfitting procedure in Figure \ref{fig-psnr-train}. As can be seen, COIN outperforms JPEG after 15k iterations and continues improving beyond that. While the optimization can be noisy, we simply save the model with the best PSNR.

\textbf{Architecture choice}. In Figure \ref{fig-arch-choice}, we show the performance of various valid architectures of size 0.3bpp. As can be seen, the quality of compression depends on the architecture choice, with different optimal architectures for different bpp values, see Appendix A for details.

\vspace{-5pt}
\section{Scope, limitations and future work}
\vspace{-5pt}
\textbf{Limitations}. The main limitation of our approach is that encoding is slow, because we have to solve an optimization problem for each encoded image.
However, paying a significant computational cost upfront to compress content for delivery to many receivers is a standard practice in the setting of one-to-many media distribution, e.g., at Netflix \citep{netflix-optim}.
Nevertheless, this limitation could likely be sidestepped with meta-learning \citep{sitzmann2020metasdf, tancik2020learned} or amortized inference \citep{trevithick2020grf,yu2020pixelnerf} approaches.
Further, at decoding time, we are required to evaluate the network at every pixel location to decode the full image. 
However, this computation can be embarrassingly parallelized to the point of a single forward pass for all pixels. 
Finally, our method performs worse than state of the art compression methods. However, we believe there are several promising directions to reduce this gap.

\textbf{Future work}. Recent work in generative modeling of implicit representations \citep{dupont2021generative} suggests that learning a distribution over the function weights could translate to significant compression gains for our approach. 
In addition, exploring meta-learning or other amortization approaches for faster encoding could be an important direction for future work \citep{sitzmann2020metasdf, tancik2020learned}. 
Refining the architectures of the functions representing the images (through neural architecture search or pruning for example) is another promising avenue. While we simply converted weights to half-precision in this paper, large gains in performance could likely be made by using more advanced model compression \citep{havasi2019minimal,vanbaalen2020bayesian,jacob2018quantization}. 
Finally, as implicit representations map arbitrary coordinates to arbitrary features \citep{tancik2020fourier, sitzmann2020implicit, dupont2021generative}, it would be interesting to apply our method to different types of data, such as video or audio.


\vspace{-5pt}
\section{Conclusion}
\vspace{-5pt}
In this paper, we proposed COIN, a new method for compressing images by fitting neural networks to pixels and storing the weights of the resulting models. We showed through experiments that this simple approach can outperform JPEG at low bit-rates, even without the use of entropy coding.
We hope that further work in this area will lead to a novel class of methods for neural data compression.

\bibliography{references,adam-refs,milad-refs}
\bibliographystyle{plain}

\newpage
\appendix

\section{Experimental details}

All models were trained using Adam for 50k iterations. We used MLPs with 2 input dimensions (corresponding to $(x,y)$ coordinates) and 3 output dimensions (corresponding to RGB values). The coordinates were normalized to lie in $[-1, 1]$ and the RGB values were normalized to lie in $[0, 1]$. We used sine non-linearities at every layer except the last and used the initialization described in \citep{sitzmann2020implicit}. 
We used a learning rate of 2e-4.
Below we describe the architectures for each bpp level.

\begin{itemize}
    \item 0.07bpp. Number of layers: 5, width of layers: 20.
    \item 0.15bpp. Number of layers: 5, width of layers: 30.
    \item 0.3bpp. Number of layers: 10, width of layers: 28.
    \item 0.6bpp. Number of layers: 10, width of layers: 40.
    \item 1.2bpp. Number of layers: 13, width of layers: 49.
\end{itemize}

The code to reproduce all experiments in the paper can be found at \url{https://github.com/EmilienDupont/coin}.


\section{Additional results}

\begin{figure}[t]
    \begin{center}
    \includegraphics[width=\textwidth]{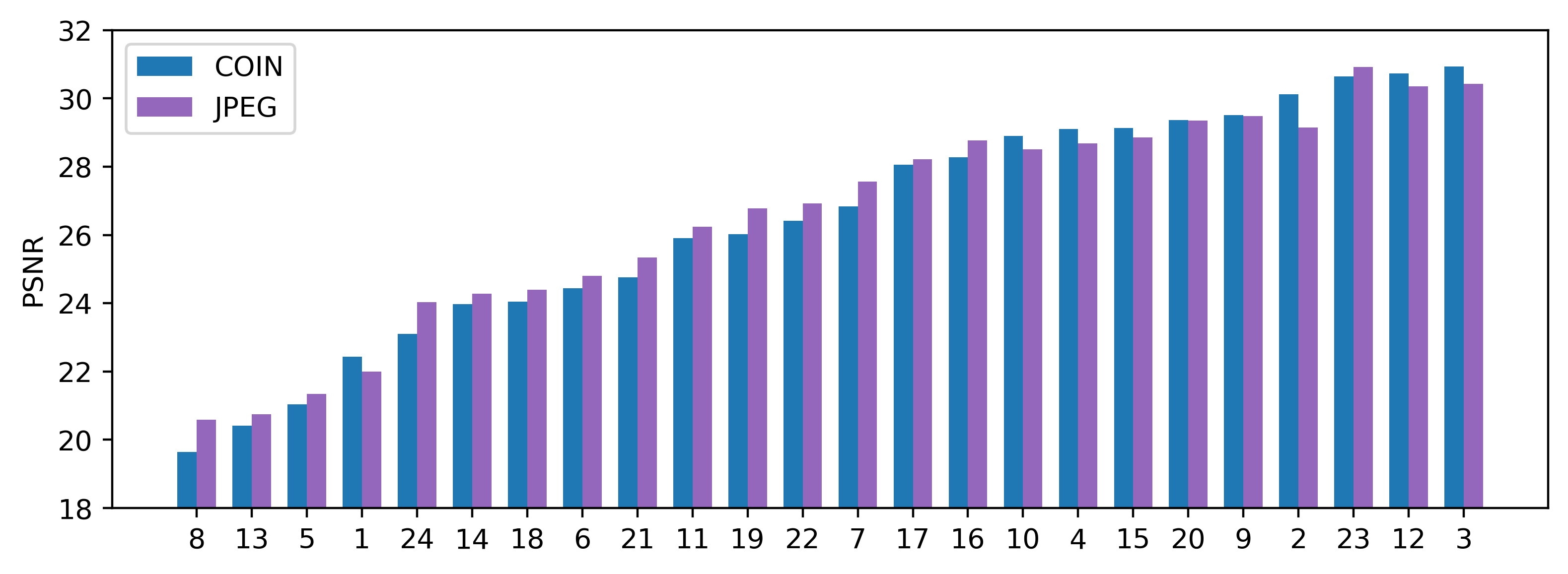}
    \end{center}
    \caption{Histogram of PSNR for all images in the Kodak dataset for COIN and JPEG.} \label{fig-psnr-hist}
\end{figure}

In Figure \ref{fig-psnr-hist}, we plot the performance of COIN and JPEG at 0.3bpp (since COIN and JPEG perform similarly at this bit-rate) for all images in the Kodak dataset. As can be seen, the distortion values closely follow each other: images that are difficult for COIN to encode are also difficult for JPEG to encode.  

\section{Qualitative results}

We include qualitative results comparing the compression artifacts from COIN and JPEG on the Kodak dataset.

\begin{figure}[t]
    \begin{center}
    \includegraphics[width=\textwidth]{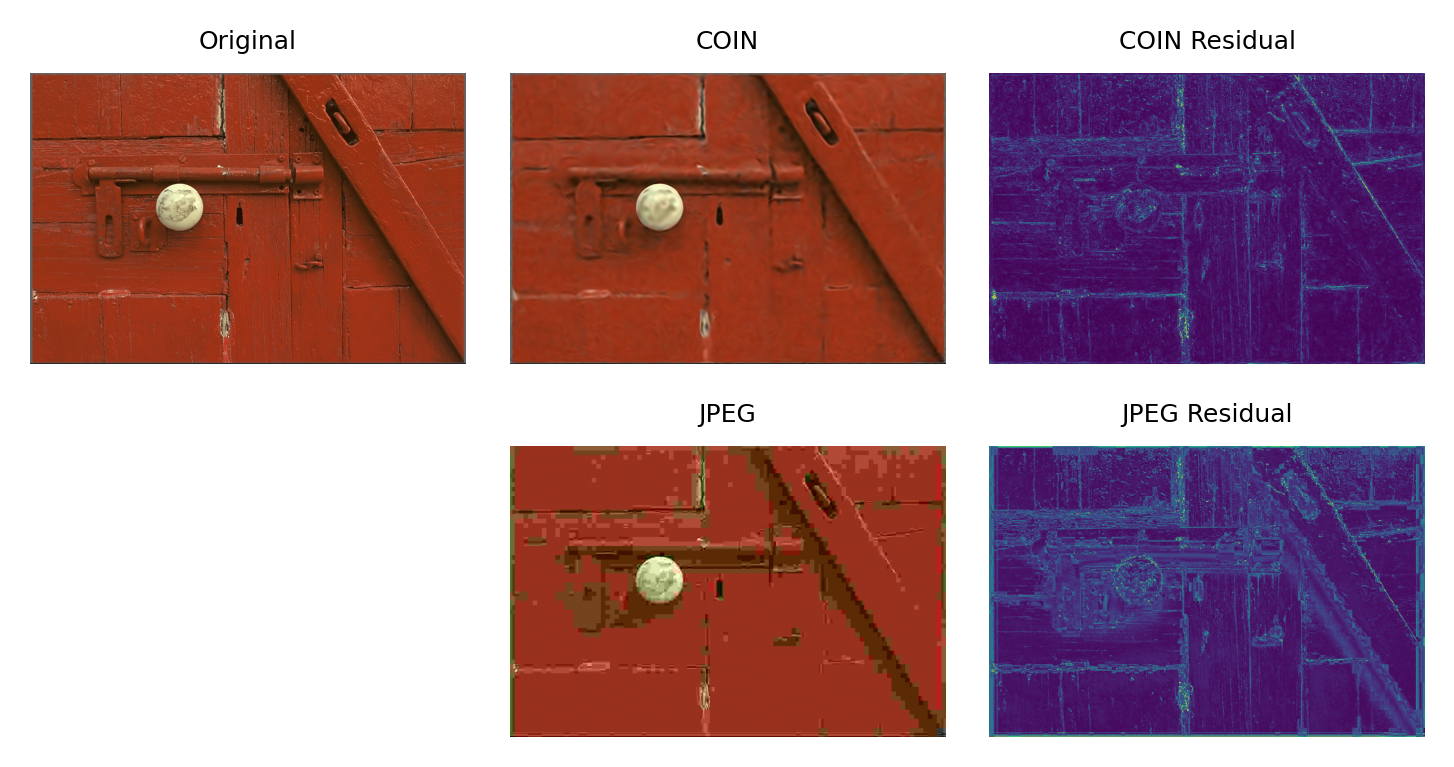}
    \end{center}
    \caption{Comparison between COIN and JPEG on image 2 at 0.15bpp. The PSNRs are 28.69dB and 24.67dB respectively.}
\end{figure}

\begin{figure}[t]
    \begin{center}
    \includegraphics[width=\textwidth]{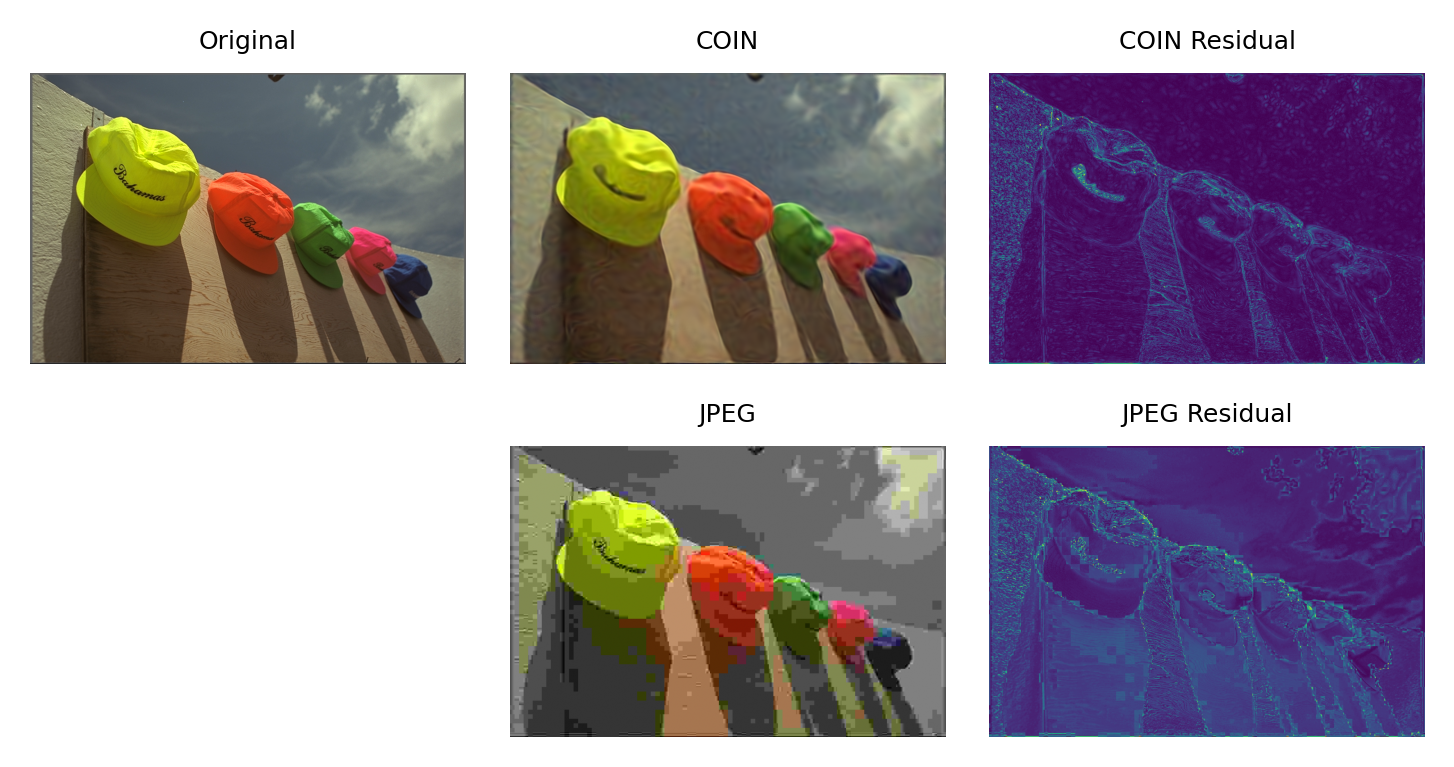}
    \end{center}
    \caption{Comparison between COIN and JPEG on image 3 at 0.15bpp. The PSNRs are 29.02dB and 23.63dB respectively.}
\end{figure}

\begin{figure}[t]
    \begin{center}
    \includegraphics[width=\textwidth]{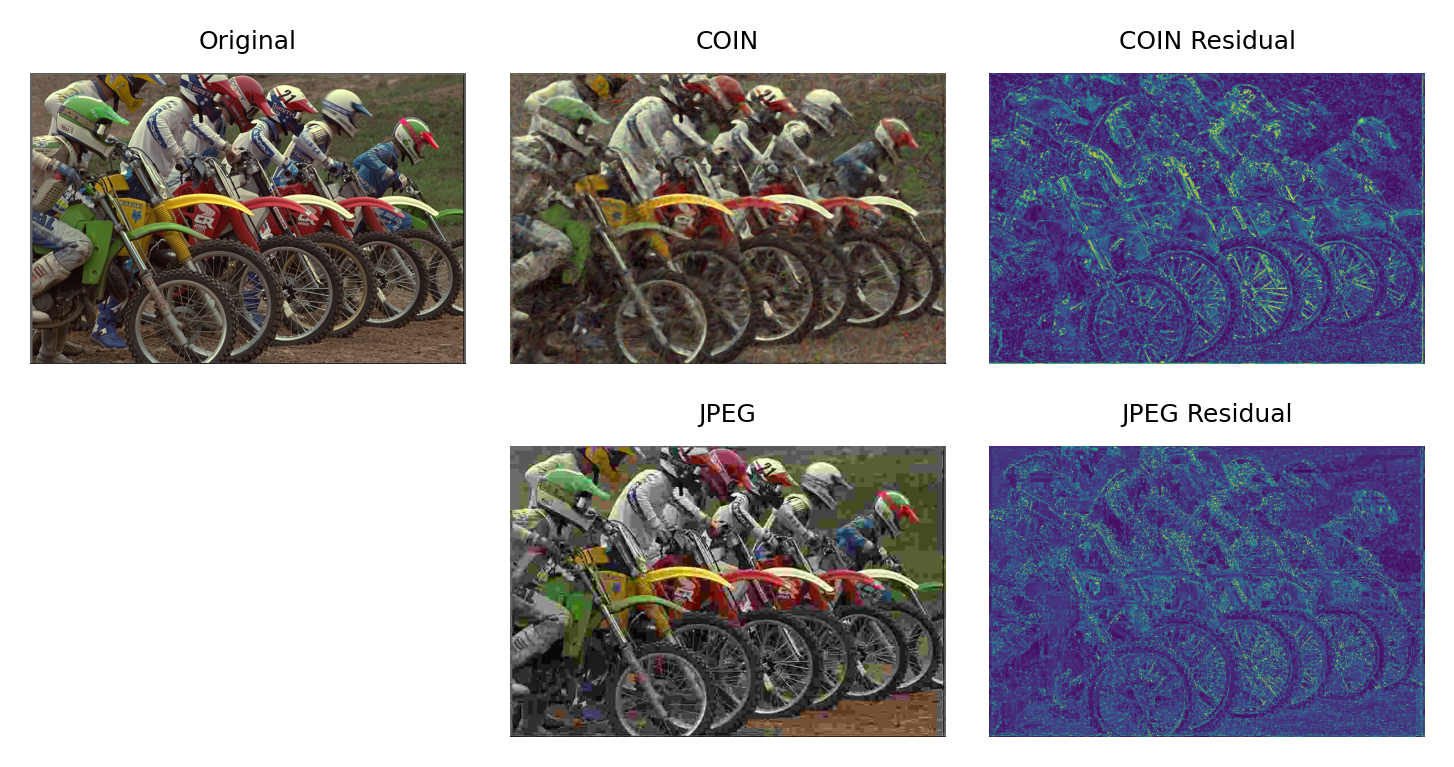}
    \end{center}
    \caption{Comparison between COIN and JPEG on image 5 at 0.3bpp. The PSNRs are 20.97dB and 21.34dB respectively.}
\end{figure}

\begin{figure}[t]
    \begin{center}
    \includegraphics[width=\textwidth]{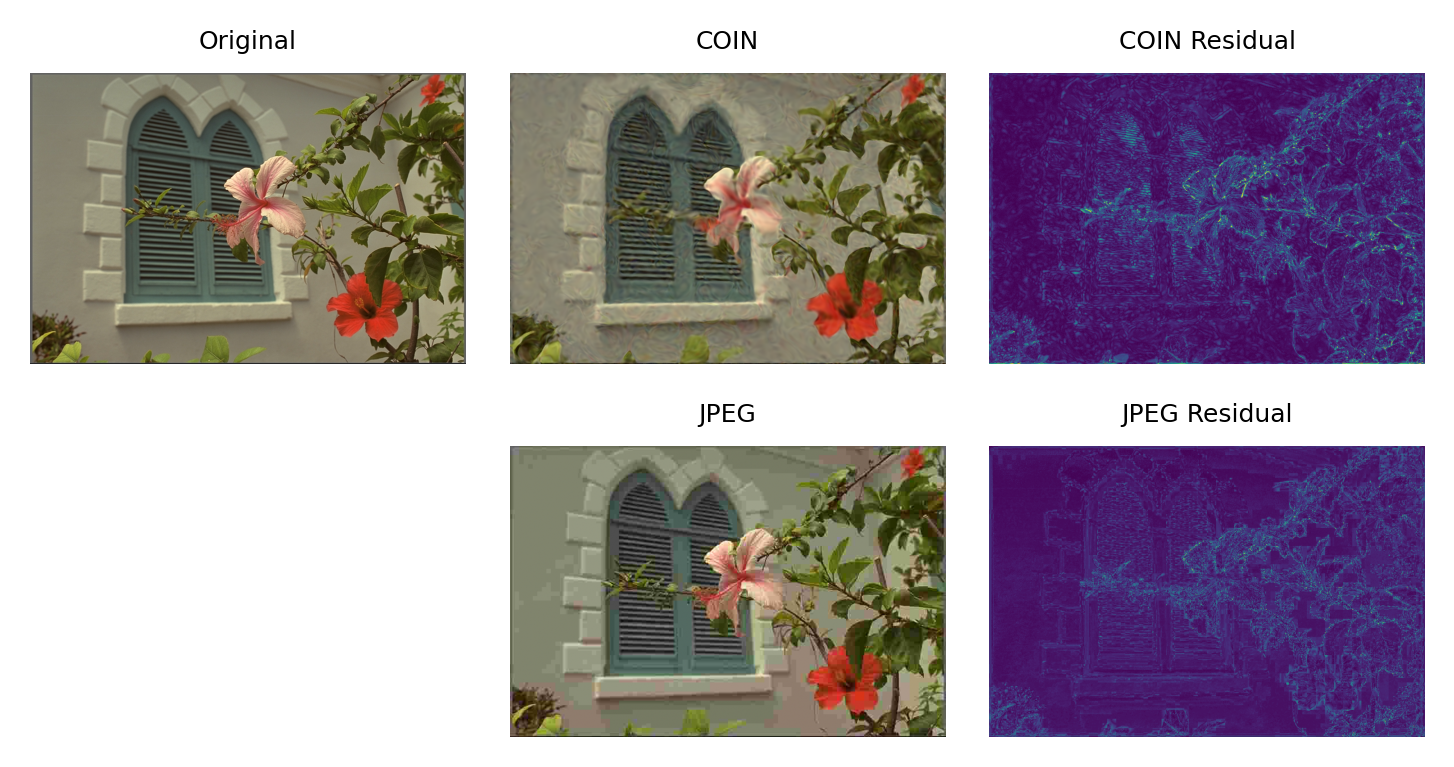}
    \end{center}
    \caption{Comparison between COIN and JPEG on image 7 at 0.3bpp. The PSNRs are 26.92dB and 27.56dB respectively.}
\end{figure}

\begin{figure}[t]
    \begin{center}
    \includegraphics[width=\textwidth]{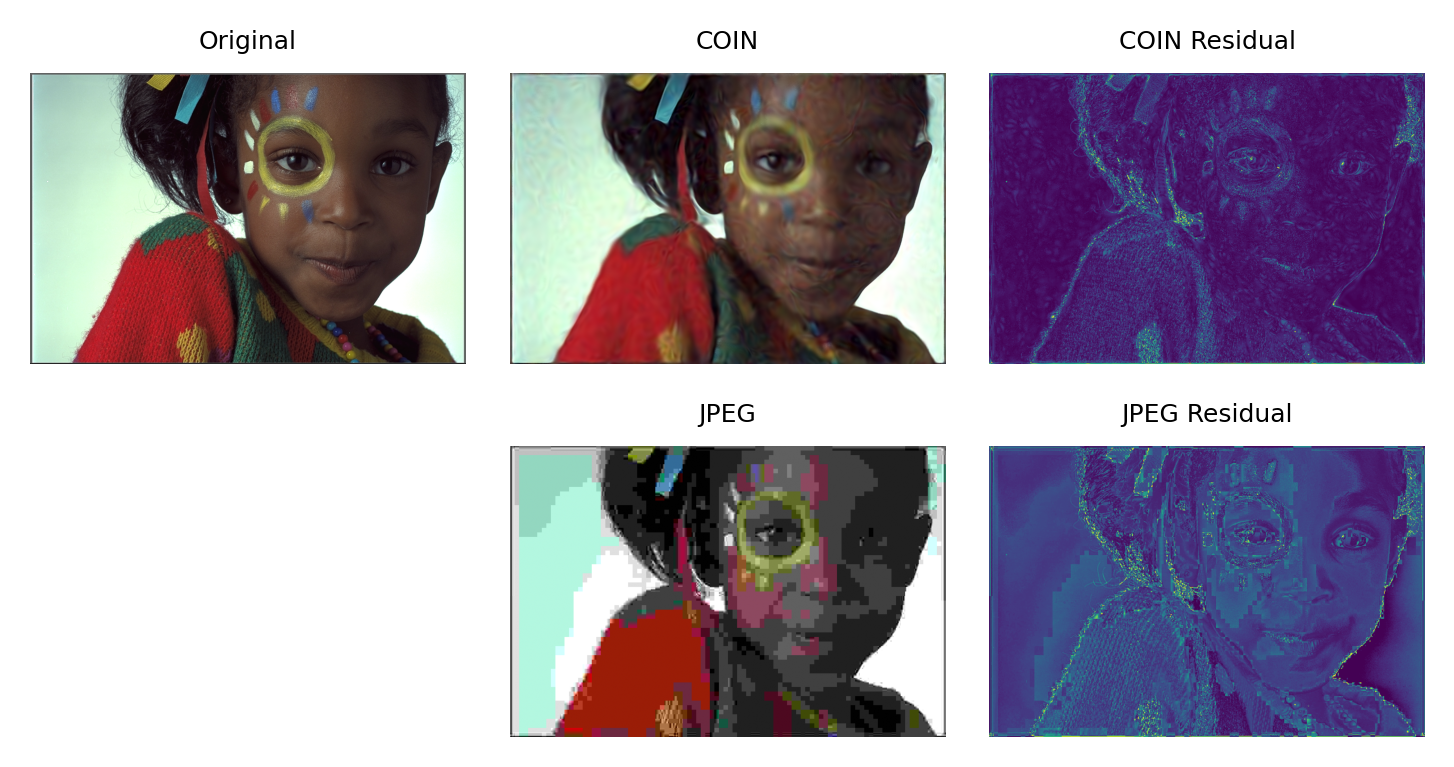}
    \end{center}
    \caption{Comparison between COIN and JPEG on image 15 at 0.15bpp. The PSNRs are 27.35dB and 21.74dB respectively.}
\end{figure}

\begin{figure}[t]
    \begin{center}
    \includegraphics[width=\textwidth]{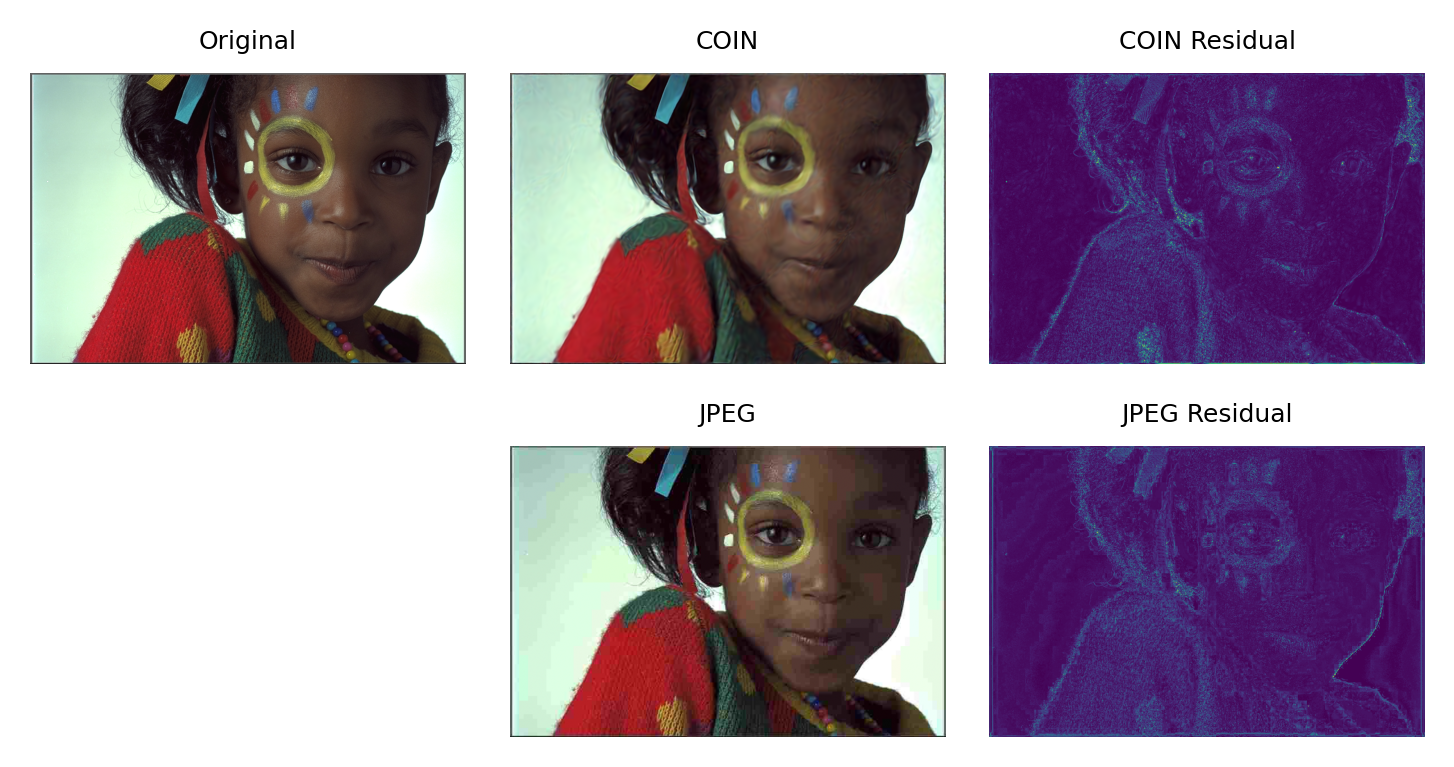}
    \end{center}
    \caption{Comparison between COIN and JPEG on image 15 at 0.3bpp. The PSNRs are 29.31dB and 28.85dB respectively.}
\end{figure}

\begin{figure}[t]
    \begin{center}
    \includegraphics[width=\textwidth]{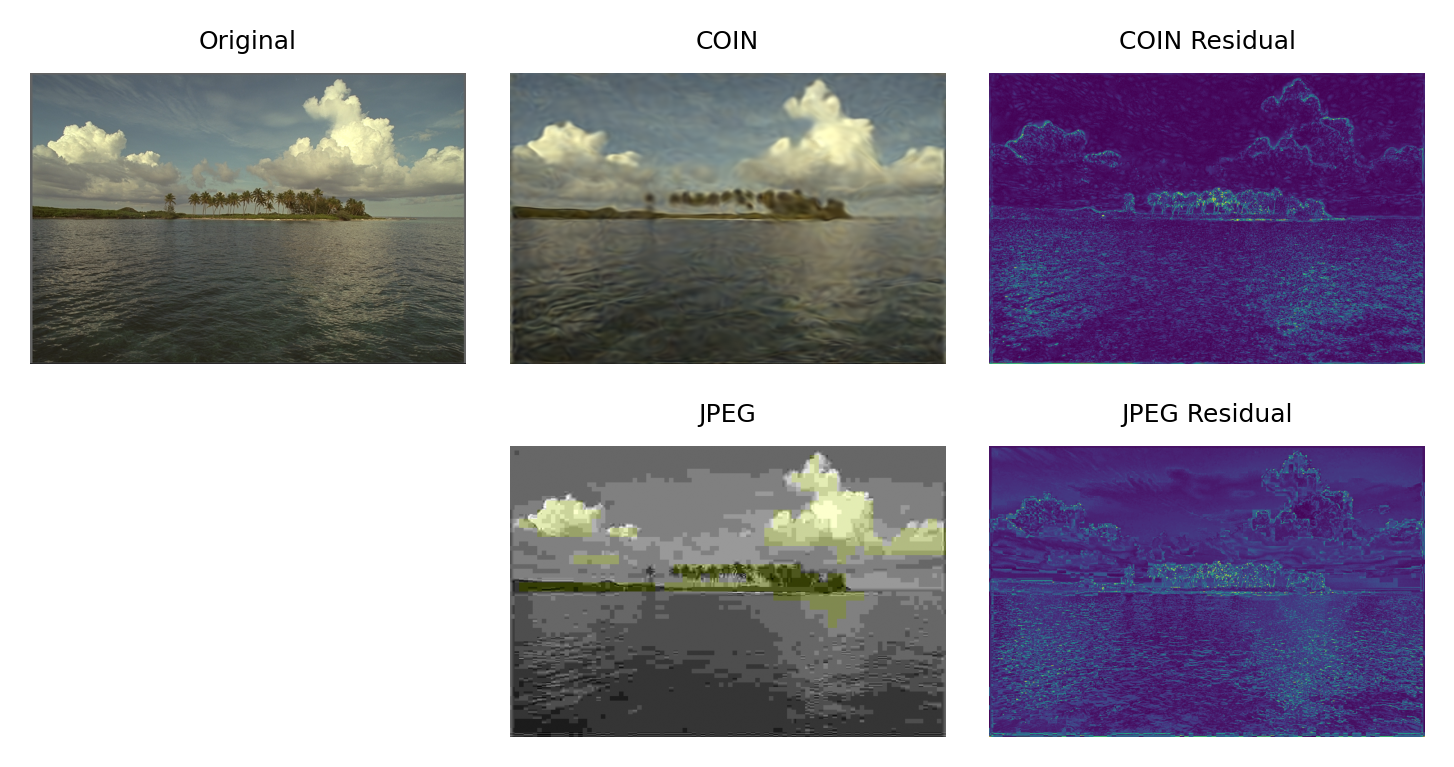}
    \end{center}
    \caption{Comparison between COIN and JPEG on image 16 at 0.15bpp. The PSNRs are 27.19dB and 24.16dB respectively.}
\end{figure}

\begin{figure}[t]
    \begin{center}
    \includegraphics[width=\textwidth]{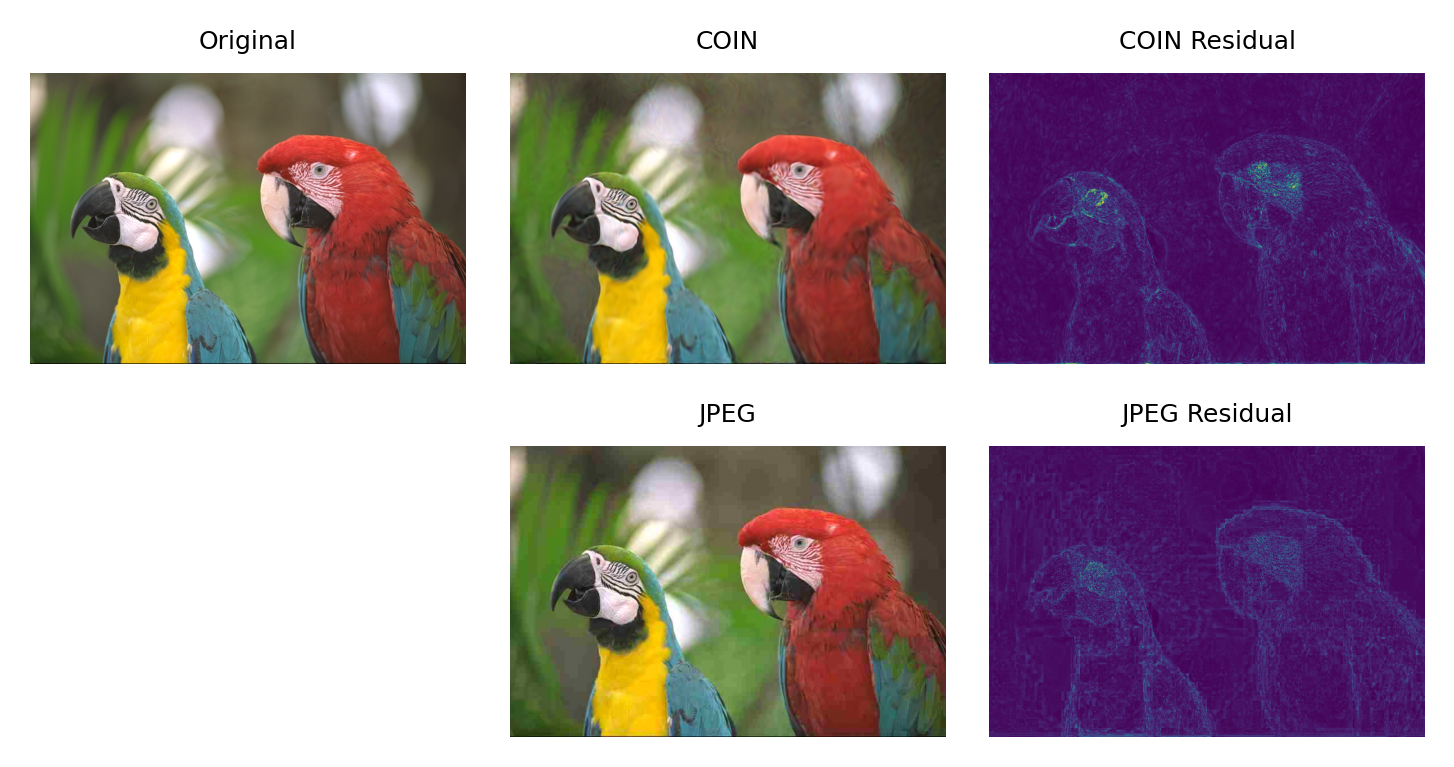}
    \end{center}
    \caption{Comparison between COIN and JPEG on image 23 at 0.3bpp. The PSNRs are 31.08dB and 30.92dB respectively.}
\end{figure}

\end{document}